\documentclass[preprint,showpacs,amsmath,amssymb]{revtex4}
\usepackage{amsmath} %

\usepackage{graphicx}
\usepackage{amsfonts}
\usepackage{dcolumn}
\usepackage{bm}

\begin{document}

\title{Nucleation pathways on complex networks}

\author{Chuansheng Shen$^{1,2}$}

\author{Hanshuang Chen$^{3}$}

\author{Zhonghuai Hou$^{1}$}\email{hzhlj@ustc.edu.cn}

\affiliation{$^{1}$Hefei National Laboratory for Physical Sciences
at Microscales, \& Department of Chemical Physics, University of
 Science and Technology of China, Hefei, 230026, China \\
 $^{2}$Department of Physics, Anqing Teachers College, Anqing, 246011, China \\
 $^3$School of Physics and Material Science, Anhui University, Hefei, 230039, China}

\date{\today}

\begin{abstract}

Identifying nucleation pathway is important for understanding the
kinetics of first-order phase transitions in natural systems. In the
present work, we study nucleation pathway of the Ising model in
homogeneous and heterogeneous networks using the forward flux
sampling method, and find that the nucleation processes represent
distinct features along pathways for different network topologies.
For homogeneous networks, there always exists a dominant nucleating
cluster to which relatively small clusters are attached gradually to
form the critical nucleus. For heterogeneous ones, many small
isolated nucleating clusters emerge at the early stage of the
nucleation process, until suddenly they form the critical nucleus
through a sharp merging process. By analyzing the properties of the
nucleating clusters along the pathway, we show that the main reason
behind the different routes is the heterogeneous character of the
underlying networks.
\end{abstract}

\maketitle

\section{Introduction}

Nucleation is a fluctuation-driven process that initiates the decay
of a metastable state into a more stable one \cite{Kashchiev2000}.
It is usually involved in first-order phase transitions and along
with growth of a new phase
\cite{RMP90000251,ARP95000489,NAT02000811}. Many important phenomena
in nature, including crystallization \cite{JCP97003634,AMI09001203},
fractures \cite{NAT91000039,PRL97003202}, glass formation
\cite{PRE98005707}, and protein folding \cite{PNAS9510869}, to list
just a few, are associated with nucleation. Despite much attention,
many aspects of nucleation processes in complex systems are still
unclear and deserve more investigation.

The Ising model is a paradigm for many phenomena in statistical
physics. It has also been widely used to study the nucleation
process. For instance, in two-dimensional lattices, Allen \emph{et
al} discovered that shear can enhance the nucleation rate and the
rate peaks at an intermediate shear rate \cite{JCP08134704}. Sear
found that a single impurity may considerably enhance the nucleation
rate \cite{JPC06004985}. Page and Sear reported that the existence
of a pore may lead to two-stage nucleation, and the overall
nucleation rate can reach a maximum level at an intermediate pore
size \cite{PRL06065701}. The nucleation pathway of the Ising model
in three-dimensional lattices has also been studied by Sear and Pan
\cite{JCP08164510,JPC0419681}. In addition, the Ising model has been
frequently used to test the validity of classical nucleation theory
(CNT)
\cite{EPJ98000571,JCP99006932,JCP00001976,PRE05031601,PRL09225703,PRE10030601,PRE10011603}.
Nevertheless, all these studies are limited to regular lattices in
Euclidean space.

 Since many real systems can be properly modeled by network-organized
structure \cite{RMP02000047,AIP02001079,SIR03000167}, it is thus an
interesting topic to explore nucleation process in complex networks.
 Very recently, our group have studied nucleation
dynamics on scale-free (SF) networks \cite{PRE11031110} and modular
networks \cite{PRE11046124}. In these two papers, we mainly focused
on the nucleation rate and system size effects. We found that, for
SF networks, the nucleation rate decays exponentially with network
size, while the critical nucleus size increases linearly. For
modular networks, as the network modularity worsens the nucleation
undergoes a transition from a two-step to one-step process and the
nucleation rate shows a nonmonotonic dependence on the modularity.
As we know, network topology could play important role in the
system's dynamics, involving not only the stationary state, but also
the detailed pathway. For instance, it was shown that network
heterogeneity could drastically influence the path to oscillator
synchronization \cite{PRL07034101}. Nevertheless, how network
topology would influence the nucleation pathway has not been
reported yet. Motivated by this, we will study the different roles
of network architectures in the formation of nucleating clusters,
which can reveal the nucleation pathways of the Ising model in the
underlying networks.

Since nucleation is an activated process, it can be extremely slow.
Therefore, direct simulations can take excessive amounts of time. To
overcome this difficulty, in the present work, we adopt the forward
flux sampling (FFS) \cite{PRL05018104} approach proposed recently,
which is efficient and easy to implement to study rare events. We
employ Erd\"{o}s-R\'{e}nyi (ER) and SF networks as the paradigm of
homogeneous and heterogeneous networks respectively. By using FFS,
we obtain lots of configurations at each interface along the
nucleation pathway. From these configurations we scrutinize and
compare the nucleating clusters in ER and SF networks. It is found
that the processes of forming the critical nucleus are qualitatively
different between the two cases of networks. For the former, a
dominant cluster arise firstly, and groups smaller clusters
gradually, while for the latter, many small clusters emerge at first
and then abruptly turn into the critical nucleus. Interestingly,
both the cluster size distributions follow power-law distributions
and the slopes are nearly the same at early nucleation stage.

The paper is organized as follows. Section \ref{sec2} presents the
details of our simulation model and the numerical methods we employ
to sampling the nucleation pathway. The numerical results are
compared for SF networks and ER ones in Sec. \ref{sec3}. A brief
summary is given in Sec. \ref{sec4}.
\section{Model and method} \label{sec2}

\subsection{ Network-organized Ising model}

We consider the Ising model on complex networks consisting of $N$
nodes. Each node is endowed with a spin variable $s_i$ that can be
either $+1$ (up) or $-1$ (down). The Hamiltonian of the system is
given by
\begin{equation}
H =  - J\sum\limits_{i < j} {A_{ij} s_i s_j } - h\sum\limits_i{s_i},
\label{eq1}
\end{equation}
where $J$ is the coupling constant and $h$ is the external magnetic
field. For convenience, we set $J=1$ in the following discussions.
The elements of the adjacency matrix of the network take $A_{ij} =
1$ if nodes $i$ and $j$ are connected and $A_{ij} =0$ otherwise. The
degree, that is the number of neighboring nodes, of node $i$ is
defined as $ k_i = \sum\nolimits_{j = 1}^N {A_{ij} }.$

The system evolves in time according to single-spin-flip dynamics
with Metropolis acceptance probabilities \cite{Lan2000}, in which we
attempt to flip each spin once, on average, during each Monte Carlo
(MC) cycle. In each attempt, a randomly chosen spin is flipped with
the probability $\min (1,e^{ - \beta \Delta E} )$, where $\beta =
1/(k_BT )$ with $k_B$ being the Boltzmann constant and $T$ the
temperature, and $\Delta E$ is the energy change due to the flipping
process. In the absence of an external magnetic field, the system
undergoes an order-disorder phase transition at the critical
temperature. Above the critical temperature, the system is
disordered where up- and down-pointing spins are roughly equally
abundant. Below the critical temperature, the system prefers to be
in either of the two states: one state with predominantly up spins,
and the other with almost down spins. In the presence of an external
field, one of these two states becomes metastable, and if initiated
predominantly in this metastable state, the system will remain for a
significantly long time before it undergoes a nucleation transition
to the thermodynamically stable state. We are interested in the
pathways for this transition.

\subsection{FFS method}

The FFS method has been successfully used to calculate rate
constants and transition paths for rare events in equilibrium and
nonequilibrium systems
\cite{JCP08134704,JPC06004985,PRL06065701,PRL05018104,JCP07114109,JCP06024102}.
For clarity, we describe the method again here, together with some
relevant details with our work. This method uses a series of
interfaces in phase space between the initial and final states to
force the system from the initial state $A$ to the final state $B$
in a ratchetlike manner. Before the simulation begins, an reaction
coordinate $\lambda$ is first defined, such that the system is in
state $A$ if $\lambda < \lambda_0$ and it is in state $B$ if
$\lambda > \lambda_M$. A series of nonintersecting interfaces
$\lambda_i$ ($0 < i <M$) lie between states $A$ and $B$, such that
any path from $A$ to $B$ must cross each interface without reaching
$\lambda_{i+1}$ before $\lambda_i$. The algorithm first runs a
long-time simulation which gives an estimate of the flux escaping
from the basin of $A$ and generates a collection of configurations
corresponding to crossings of interface $\lambda_0$. The next step
is to choose a configuration from this collection at random and use
it to initiate a trial run which is continued until it either
reaches $\lambda_1$ or returns to $\lambda_0$. If $\lambda_1$ is
reached, the configuration of the end point of the trial run is
stored. This process is repeated, step by step, until $\lambda_M$ is
reached. For more detailed descriptions of the FFS method please see
Ref. \cite{JPH09463102}.

In this work, we will use FFS to study nucleation pathways of the
equilibrium phase from the metastable spin phase. Specifically, we
let $h>0$ and start from an initial state with $s_i=-1$ for most of
the spins. We define the order parameter $\lambda$ as the total
number of up spins in the network. The spacing between adjacent
interfaces is fixed at 3 up spins. We perform $1000$ trials per
interface for each FFS sampling, from which at least $200$
configurations are saved in order to investigate the statistical
properties along the nucleation pathway. The results are obtained by
averaging over $10$ independent FFS samplings and $50$ different
network realizations.

\section{Numerical Results}  \label{sec3}

In what follows, we employ a Barab\'{a}si-Albert SF network, whose
degree distribution follows a power law $P(k) \sim k^{-\gamma}$ with
the scaling exponent $\gamma = 3$ \cite{SCI99000509}, and the
well-known ER random network.

\begin{figure}[h]
\centerline{\includegraphics*[width=0.95\columnwidth]{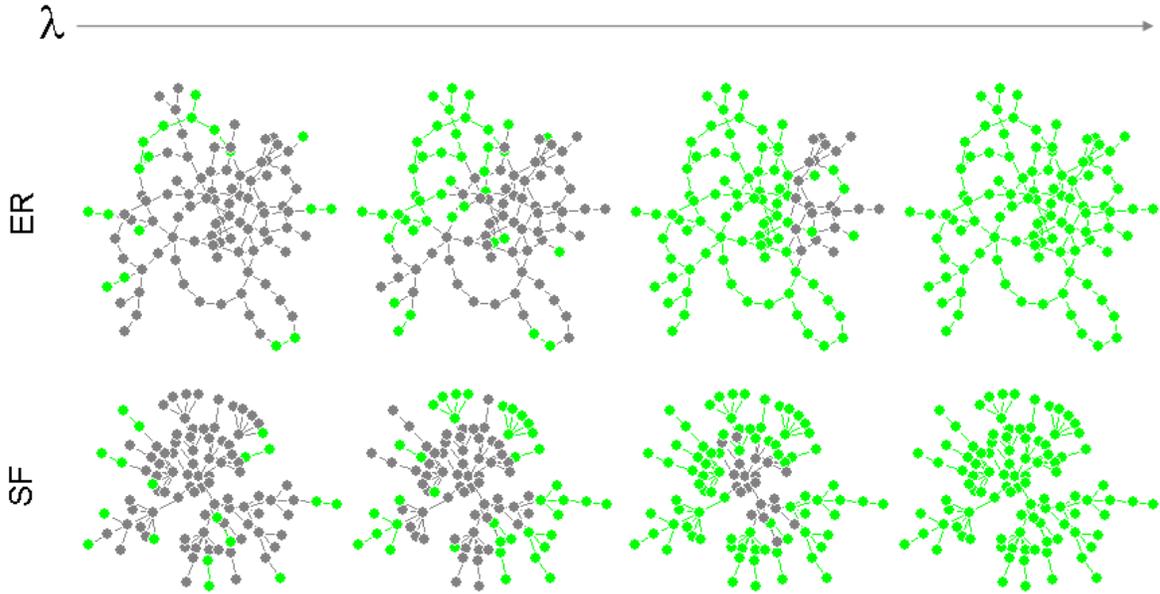}}
\caption{(Color online) Typical nucleating clusters for several
values of $\lambda $ for the two different topologies studied (ER
and SF). These shown networks contain 100 nodes, in order to have a
sizeable picture of the system. Up spins and down spins are
indicated by green circles and black circles, respectively.
\label{fig1}}
\end{figure}

In Fig. \ref{fig1}, we present schematically the evolution of local
nucleating clusters in ER and SF networks at different stages (for
clarity only show 100 nodes). Here, a nucleating cluster is defined
as the component of connected nodes with up spins. Qualitatively, it
shows distinct features along nucleation stages. In the ER case,
there always exists a dominant cluster, which groups smaller ones
gradually. While for SF networks, no dominant cluster appears at the
early stage, but then a giant cluster emerges suddenly. This
demonstrates that nucleation follows different pathways on ER and SF
networks, indicating that heterogeneity of the network topology may
play an important role.

To further elucidate detailed characteristics along the nucleation
pathway, we use FFS to generate configurations and perform detailed analysis
 on the nucleating clusters, including the largest cluster size, average
degree of the cluster nodes, the number of clusters and cluster size
distribution. According to CNT, there exists a critical nucleus size
$\lambda_c$ of the new phase, above which the system grows rapidly
to the new phase. Herein, we mainly focus on the nucleation stage
where $\lambda<\lambda_c$. In our simulation, we determine
$\lambda_c$ by computation of the committor probability $P_B$, which
is the probability of reaching the thermodynamic stable state before
returning to the metastable state. As commonly reported in the
literature \cite{JPC0419681,PRE10030601}, the critical nucleus
appears at $P_B(\lambda_c)=0.5$. Since $\lambda_c$ are different for
different networks, we thus introduce $\lambda/\lambda_c$ as the
control parameter.

For consistent comparison, we introduce $S_{max}$ as the ratio of
the size of the largest nucleating cluster to over the total number
of up spins, and plot $S_{max}$ (averaged over the ensemble at each
interface) as functions of $\lambda /\lambda _c $ in Fig.2. Clearly,
one can see that $S_{max}$ for ER networks is always larger than
that for SF ones. Specifically, at $\lambda /\lambda _c=0.5 $,
$S_{max}$ is already more than $70\% $ for ER networks, while it is
only about $30\% $ for SF ones, as shown by the dashed gray lines in
Fig.2. But when $\lambda /\lambda _c=1$ they almost tend to $100\%$
together.

\begin{figure}
\centerline{\includegraphics*[width=0.6\columnwidth]{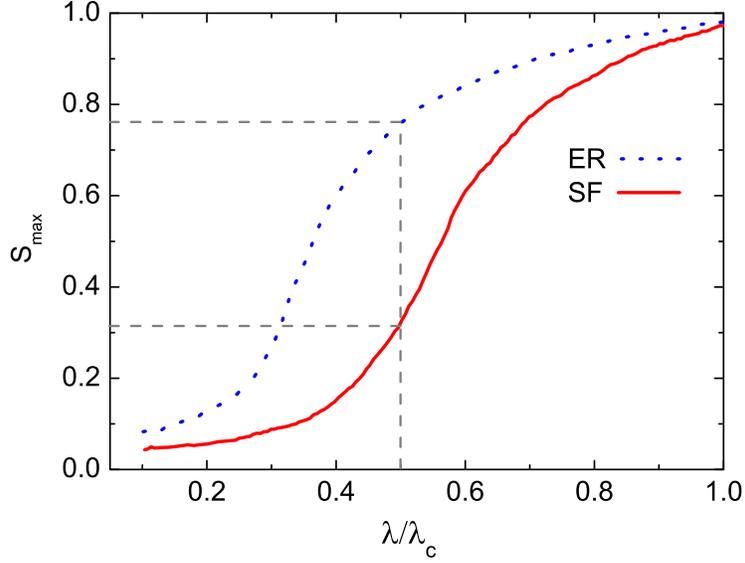}}
\caption{(Color online) The relative size $S_{max}$ of the largest
cluster as functions of $\lambda /\lambda _c $. Parameters are $N =
1000$, the average network degree $K=6$, $h=0.5$, $T/T_c=0.3$,
$\lambda_0 = 130$ and $\lambda_M = 880$. \label{fig2}}
\end{figure}

\begin{figure}
\centerline{\includegraphics*[width=0.6\columnwidth]{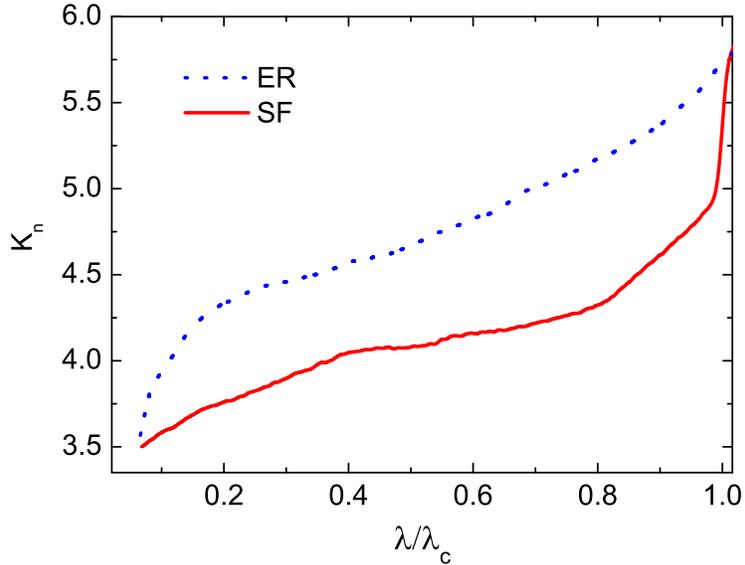}}
\caption{(Color online) Average degree $K_n$ of new phase nodes as
functions of $\lambda /\lambda _c $. Other parameters are the same
as in Fig.2. \label{fig3}}
\end{figure}

To show our results more explicitly, we investigate the average
degree $K_n$ of the nodes inside the nucleating clusters, and plot
$K_n$ as functions of $\lambda /\lambda_c$ in Fig. \ref{fig3}. As
shown, $K_n$ increases monotonically with $\lambda /\lambda _c $ for
both ER and SF networks, which means the new phase tends to grow
from those nodes with smaller degrees. For ER networks, $K_n$ grows
fast at the very beginning following by a relatively slow
increasing. For SF networks, $K_n$ increases slowly at first and
jumps sharply when approaching the critical nucleus. Such a scenario
is consistent with Fig.\ref{fig1} and Fig.\ref{fig2}.

\begin{figure}
\centerline{\includegraphics*[width=0.6\columnwidth]{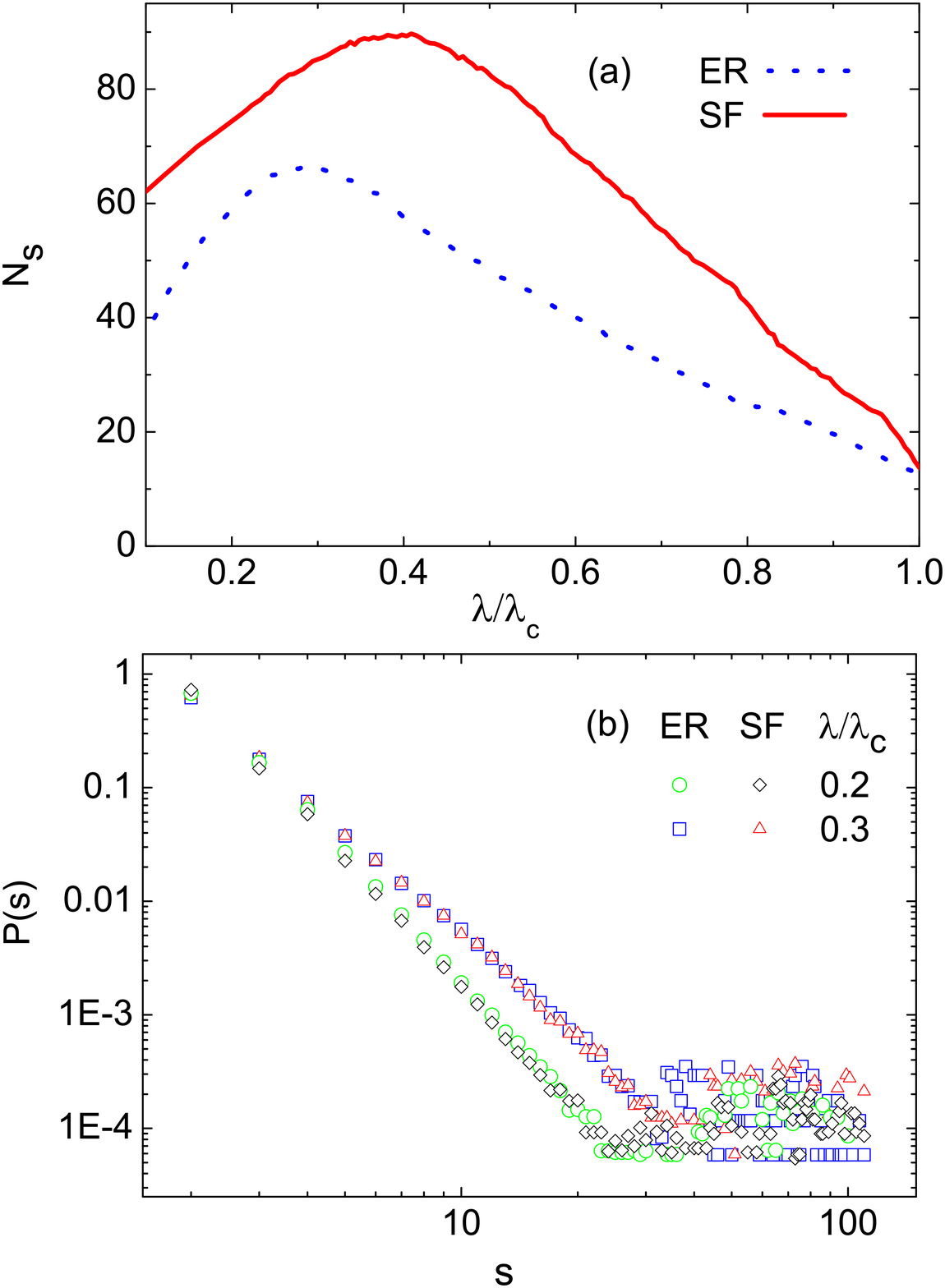}}
\caption{(Color online) (a) The number $N_s$ of nucleating clusters
as functions of $\lambda /\lambda _c $. (b) Size distribution $P(s)$
of nucleating clusters, on a log-log scale, $s$ denote the size of
nucleating clusters. Other parameters are the same as in Fig.2.
\label{fig4}}
\end{figure}

To better understand the aforementioned differences, we present the
number $N_s$ of the nucleating clusters as functions of $\lambda
/\lambda _c $ in Fig. \ref{fig4}(a). We observe that $N_s$
non-monotonically depends on $\lambda /\lambda _c $ and the numbers
of clusters in SF networks are always more than that in ER ones. On
the other hand, $N_s$ for both networks approach the same magnitude
near the formation of critical nucleus, but it decreases much more
sharply in SF networks which is also consistent with the picture
shown in Fig.\ref{fig1} to \ref{fig3}. In Fig. \ref{fig4}(b), the
cluster size distributions $P(s)$ for $\lambda/\lambda_c=0.2$ and
$0.3$ are shown. Interestingly, $P(s)$ follow apparent power-law
distributions in the small size range for both types of networks,
and in addition, the exponents are nearly the same for fixed
$\lambda/\lambda_c$. The power law breaks in the large size range,
where large clusters dominate.

The above results can be qualitatively understood in terms of CNT.
CNT assumes that the formation of a nucleus lies in two competing
factors: the bulk energy gain of creating a new up spin which favors
the growth of the nucleus, and the interfacial energy cost, an
opposing factor, which is due to the creation of new boundary links
between up and down spins. That is, the change of free energy
$\bigtriangleup F$ may be written as $\bigtriangleup
F(\lambda)=-2h\lambda+\sigma \lambda$, where $\sigma$ denotes the
effective interfacial free energy, which mainly depends on the
average number of boundary links that an up-spin node has.
Obviously, a node with more boundary links is more difficult to
change its spin state. For SF networks, it is thus always easier for
the leaf nodes with small degrees to change state than the hubs with
large degrees. Since the degree distribution follows power-law,
there exist a lot of hubs with intermediate degrees, as well as a
few hubs with very large degrees. Usually, many leaf nodes are
connected to relatively small hubs, which are further connected to
large hubs. Therefore, only small nucleating clusters, consisted of
leaf nodes and small hubs, can form at the early stage of the
nucleation process. These small clusters are either away from each
other on the network or separated by those crucial hubs with very
large degrees. In the final stage of the nucleation, once the
crucial hubs connecting these small clusters change their states, a
giant nucleation cluster will emerge abruptly. This picture is
consistent with those results shown in the above figures. For ER
networks, however, the degree distribution follows Poisson
distribution and no crucial hub exists, such that those new-formed
clusters are usually connected together and one would not expect a
sharp increase in the cluster size, which is observed in SF
networks.

\section{Conclusions} \label{sec4}

In summary, we have studied nucleation pathways of the Ising model
with Metropolis spin-flip dynamics in ER and SF networks using the
FFS method. Concerning the former, there always exists a dominant
cluster which groups small clusters gradually until the critical
nucleus is formed; while concerning the latter, many isolated small
clusters grow separately which suddenly merge together into the
critical nucleus. We have performed detailed analysis involving the
nucleating clusters along the nucleation pathway, including the
cluster size as well as its distribution, the mean degree inside the
cluster, and so on, to further demonstrate the above scenario. The
distinct nucleation pathways between ER and SF networks further
emphasize the very role of network topology. Our study may provide a
valuable understanding of how first-order phase transitions take
place on complex networks, which could be of great importance not
only for physical systems, but also for social and biological
networks.

\begin{acknowledgments}
This work was supported by the National Natural Science Foundation
of China (Grant Nos.21125313, 20933006 and 91027012). C.S.S. was
also supported by the Key Research Foundation of Higher Education of
Anhui Province of China (Grant No.KJ2012A189).
\end{acknowledgments}

%
\bibliographystyle{apsrev}

\end{document}